\documentclass[prb,twocolumn,showpacs,superscriptaddress]{revtex4}
\usepackage{graphicx}
\usepackage{dcolumn}% Align table columns on decimal point
\usepackage{bm}% bold math

\begin{document}
\title{Charge instabilities in strongly correlated bilayer systems}
\author{G. Seibold}
\affiliation{Institut f\"ur Physik, BTU Cottbus, PBox 101344, 
         03013 Cottbus, Germany}
\date{\today}
\begin{abstract}
We investigate the charge-instabilities of the 
Hubbard-Holstein model with two coupled layers.
In this system the scattering processes naturally separate into 
contributions which are either
symmetric or antisymmetric combinations with respect to exchange of the
layers. It turns out that the short-range strong correlations suppress
finite wave-vector nesting instabilities for both symmetries 
but favor the occurrence of
phase separation in the symmetric channel.
Inclusion of a sizeable long-range Coulomb (LRC) interaction 
frustrates the $q=0$
instabilities and supports the formation of incommensurate charge-density
waves (CDW). Upon reducing doping from half-filling and for small 
electron-phonon coupling $g$  
the CDW instability first occurs  in the antisymmetric channel but
both instability lines merge with increasing g. 
While LRC forces always suppress the phase separation instability in the 
symmetric channel, the CDW period in the antisymmetric sector tends to 
infinity ($q_c\to 0$) for sufficiently small Coulomb interaction.
This feature allows for the possibility of singular scattering
over the whole Fermi surface.
We discuss possible implications of our results 
for the bilayer high-T$_c$ cuprates.
\end{abstract}


\maketitle
\section{Introduction}
Among the variety of cuprate superconductors most  
angle-resolved photoemission spectroscopy (ARPES) experiments
have been made on the bilayer compound Bi$_2$Sr$_2$CaCu$_2$O$_{8+\delta}$ 
(Bi2212) due to the advantage of good cleavage planes and its
nearly perfectly two-dimensional electronic structure.~\cite{CAMP0}
However, only recent improvements in the resolution of ARPES
measurements allowed for the detection of the band splitting 
due to coherent c-axis coupling of the two layers within a unit cell
.~\cite{FENG,CHUANG} 
The same feature has also been observed in modulation-free
(Bi,Pb)-2212 \cite{KORDYUK} and there is
experimental evidence that the magnitude of the bilayer splitting 
is constant over a
large range of doping.~\cite{CHUANG2}
The splitting obeys the expected symmetry of LDA computations
\cite{ANDERSEN,LIECHTENSTEIN} being essentially zero along the 
diagonals $(0,0)\to(\pi,\pi)$ and (in the normal state) 
acquires a value between 88meV \cite{FENG} and 110meV \cite{CHUANG}
at the $(\pi,0)$ point of the Brillouin zone (BZ). 

Below T$_c$, the ARPES spectra display additional features which 
can be interpreted in terms of a coupling of the charge carriers
to a collective mode. As a consequence the dispersion along
the nodal direction shows a break at some characteristic energy $\omega_0$
\cite{LANZARA} with an increased effective mass for binding energies smaller
than $\omega_0$. It has been shown that an analogous anomaly
is also present in the bonding band dispersion of Bi2212 around $(\pi,0)$
(M-point) which reveals as an additional peak-dip hump feature in
the ARPES line shape.~\cite{GROMKO} However, more recent ARPES experiments on 
underdoped (Bi,Pb)-2212 \cite{KIM} have revealed a similar mass
renormalization in bonding and antibonding band. 

Concerning the physical origin of these mode-type features the various
proposals include a magnetic resonance 
(see e.g. Refs. \onlinecite{CAMP,ESCHRIG1,ESCHRIG2}), the
coupling to phonons \cite{LANZARA} or incommensurate
charge-density waves (ICDW) \cite{GOETZ,VARL} as the source of the
associated scattering.
Further on from neutron scattering experiments 
(see e.g. Ref. \onlinecite{TRAN}) it is
known that the spin fluctuations between the layers 
in YBa$_2$CU$_3$O$_{7-\delta}$ 
are antiferromagnetically correlated so that the corresponding
exchange potential is antisymmetric. 
However, since the weight of the resonance is quite small, large
coupling constants are required in order to reproduce the spectral
features within a magnetic mechanism and therefore this scenario
is still controversially discussed.~\cite{KIV}

In this paper we show that antisymmetric scattering in a bilayer system is 
not unique to a magnetic interaction but may also occur in the 
charge sector close to a ICDW instability.
Our analysis can be viewed as an extension of the
theory proposed by Castellani et al. in Ref. \onlinecite{CAST} according
to which the anomalous electronic properties of high-T$_c$ materials  
are determined by a quantum critical point (QCP) located near optimal doping.
The occurence of such an instability towards ICDW formation can be
theoretically substantiated by considering the interplay between
phase separation (PS) and the long-range Coulomb interaction.
PS is a natural feature of systems where strong electronic correlations
lead to a substantial reduction of the kinetic energy. 
As a consequence short-range range attractive interactions (e.g.
of phononic or magnetic origin) may dominate and induce a charge
aggregation in highly doped metallic regions and a simultaneous
charge depletion in spatially separated regions.
It was pointed out in Ref. \onlinecite{EMERY} that long-range Coulomb
forces oppose the charge separation suppressing long-wavelength
density fluctuations. It has been shown \cite{CAST,BECCA} that this 
'frustration' of PS may result
in a finite-momentum instability corresponding to a ICDW quantum
critical point.
Near this instability the associated singular scattering favors
the occurrence of d-wave superconductivity \cite{PERALI} and can explain
the anomalous hump-type absorption in the optical conductivity of 
overdoped cuprates.~\cite{CAPRARA}
In addition, the strong fluctuations associated with the proximity to a 
QCP may account for the dependence of the pseudogap temperature
on the characteristic time scale of the particular 
experiment.~\cite{ANDERGASSEN}

Further experimental evidence for the existence of a QCP in the phase diagram
of high-T$_c$ cuprates has been growing over the last few years
(see e.g. Ref. \onlinecite{EVIDENCE} and references therein). The idea that
the associated order in the underdoped regime is compatible with ICDW 
formation is also supported by recent scanning tunneling microscopy
(STM) experiments.~\cite{HOWALD} These measurements have revealed
the existence of a non-dispersing peak in the Fourier transformed local
density of states in slightly
overdoped Bi2212 and thus the presence of static charge order
in this compound (for a more detailed discussion on the detection
of charge order in cuprates see Ref. \onlinecite{KIV2}). 

Previous investigations of the ICDW-QCP scenario \cite{CAST,BECCA} have
been based on the two-dimensional electronic structure of 
a single CuO$_2$ plane. However, real cuprate compounds can be
prepared with a variable number of CuO$_2$ layers per unit cell which
additionally are electronically coupled along the c-axis.
Such a layered structure has profound consequences on the momentum 
dependence of the long-range Coulomb interaction  and 
on the spectrum of low-energy collective modes.
In this paper we study the simplest extension of the single-layer
case, namely we investigate possible charge instabilities in
a system consisting of two coupled layers.
In Sec. II we introduce the Hubbard-Holstein bilayer model and 
outline the evaluation of the relevant effective interactions
between quasiparticles within an $1/N$ expansion.
In a bilayer system these interactions can be either symmetric
or antisymmetric with respect to exchange of the layers.
We show in Sec. IIIa that in the absence of long-range interactions
and similar to the single-layer system the strong local repulsion
favors the occurence of a phase separation instability
in the symmetric sector. 
Long-range Coulomb forces are introduced in Sec. IIIb. It turns out
that in this case the preferred charge-instability occurs in the 
antisymmetric channel although the symmetric instability line can be rather
close. 
We finally discuss possible implications of our results for
the high-T$_c$ cuprates in Sec. IV. 
Since this paper extends previous calculations for single-layer
systems the reader is recommended to study Ref. \onlinecite{BECCA}
for more details on the subject.

\section{Formalism}
\subsection{The Model}
Starting point Hamiltonian is the Hubbard-Holstein  model
for two coupled layers: 
\begin{eqnarray}
H & = & \sum_{ij,\sigma,\alpha}t_{ij}
f^\dagger_{i\sigma,\alpha} f_{j\sigma,\alpha}+\sum_{ij, \sigma}t^\perp_{ij}
\left( f^\dagger_{i\sigma,1} f_{j\sigma,2}+ h.c.\right)\nonumber \\
&+& U\sum_i n_{i\uparrow}n_{i\downarrow} 
-\mu_0\sum_{i,\alpha}n_{i\sigma,\alpha}
+ \omega_0 \sum_{i,\alpha} A^\dagger_{i,\alpha} A_{i,\alpha}
\nonumber \\
&-& g \sum_{i\sigma,\alpha}
\left( A^\dagger_{i,\alpha}+A_{i,\alpha}\right) \left(
n_{i\sigma,\alpha} - \langle n_{i\sigma,\alpha} \rangle
\right)
\label{HHol}
\end{eqnarray}
where $f^{(\dagger)}_{i\sigma,\alpha}$ annihilates (creates) an
electron at site $R_i$ of layer $\alpha=1,2$ and $n_{i\sigma,\alpha}
=f^{\dagger}_{i\sigma,\alpha}f_{i\sigma,\alpha}$.
Note that Eq. (\ref{HHol}) does not contain long-range forces which will 
be included in Sec. IIIb.

The chemical potential is denoted by $\mu_0$ and $t_{ij}$ and $t^\perp_{ij}$
are hopping amplitudes in- and between the layers, respectively.
In the following we take as the Fourier transformed of the interlayer
hopping
\begin{equation} \label{tperp}
t^\perp(k)=t^\perp_0
\left\lbrack \left(\cos(k_x)-\cos(k_y\right))^2/4\right\rbrack
\end{equation}
motivated by LDA calculations \cite{FENG,CHUANG} and ARPES
experiments \cite{LIECHTENSTEIN} for high-T$_c$ bilayer compounds.

The operators $A_i^{(\dagger)}$ describe dispersionless phonons 
(frequency $\omega_0$) interacting with the electrons via a
local (Holstein-type) coupling.   
Note that the electron-phonon coupling vanishes on the mean-field level 
since it is incorporated only via the density fluctations \cite{COMM}. 

Furtheron we take the limit $U\to \infty$ which can be considered
within a standard slave-boson technique.~\cite{SLABOS,RADGAUGE}
In order to implement the constraint of no double occupation the original
fermion operators are decomposed as
$f^{\dagger}_{i\sigma,\alpha}\rightarrow c^{\dagger}_{i
\sigma,\alpha}b_{i\alpha}, \,\,\, {f}_{i\sigma,\alpha}
\rightarrow b^{\dagger}_{i\alpha} c_{i\sigma,\alpha}$.
Moreover it is convenient to introduce the limit of large orbital degeneracy
N to introduce a small parameter $1/N$ for a perturbative expansion.
The new fermion and boson operators are related by the constraint
\begin{equation}
\sum_{\sigma}c_{i\sigma\alpha}^{\dagger}c_{i\sigma\alpha}+b_{i\alpha}^{\dagger}
b_{i\alpha}=N/2
\end{equation}
which is implemented below by introducing an additional
local Lagrange multiplier $\lambda_{i\alpha}$.
Within the large N expansion the model can then be represented as a
functional integral
\begin{eqnarray}
 Z & = & \int Dc^{\dagger}_{\sigma}Dc_{\sigma} Db^{\dagger}Db
D\lambda DA DA^\dagger e^{-\int_0^\beta Sd\tau}, \label{funcint}\\
  S & = & \sum_i \left[ \sum_{\sigma} c^{\dagger}_{i \sigma}
 {{\partial c_{i\sigma}} \over {\partial \tau}}
 +b^{\dagger}_{i} {{\partial b_{i}} \over {\partial \tau}}
+A^{\dagger}_{i} {{\partial A_{i}} \over
{\partial \tau}} \right] + H \label{action}
\end{eqnarray}
with
\begin{eqnarray}
H & = & \frac{1}{N} \sum_{ij,\sigma,\alpha}t_{ij}
c^\dagger_{i\sigma,\alpha} c_{j\sigma,\alpha}b^\dagger_{j,\alpha}
b_{i,\alpha}
 + \sum_{i\sigma,\alpha}(-\mu_0+i\lambda_{i,\alpha}) n_{i\sigma,\alpha}
\nonumber\\
&+&\frac{1}{N}\sum_{ij, \sigma}t^\perp_{ij}
\left( c^\dagger_{i\sigma,1} c_{j\sigma,2}b^\dagger_{j,2}
b_{i,1} + H.c.\right)
+ \omega_0 \sum_{i,\alpha} A^\dagger_{i,\alpha} A_{i,\alpha}\nonumber\\
&+&  \sum_{i,\alpha}  i\lambda_{i,\alpha}
\left( b^{\dagger}_{i,\alpha} b_{i,\alpha}
-{N\over 2}\right)\nonumber\\
&-& \frac{g}{\sqrt{N}} \sum_{i\sigma,\alpha}
\left( A^\dagger_{i,\alpha}+A_{i,\alpha}\right) \left(
n_{i\sigma,\alpha} - \langle n_{i\sigma,\alpha} \rangle
\right)
\label{HHHam}
\end{eqnarray}
where we have rescaled the hopping $t^{(\perp)}_{ij}\to t^{(\perp)}_{ij}/N$
and the el.-ph. coupling constant $g\to g/\sqrt{N}$ in order
to compensate for the presence of N fermionic degrees of freedom per site.  
The average number of particle per
cell and plane is $n=n_1=n_2= (1-\delta)N/2$ and $\delta=0$ corresponds to
half-filling, when one half electron per cell and per spin
flavor is present in the system.

The mean-field self-consistency equations
are obtained by requiring the stationarity of the mean-field
free energy and they determine the values of
$\langle b_{i,1}\rangle^2=\langle b_{i,2}\rangle^2
=b_0^2\equiv N r_0^2$
and of $\lambda_0\equiv \langle \lambda_{i,1} \rangle
=\langle \lambda_{i,1} \rangle$. Then the
mean-field Hamiltonian reads
\begin{eqnarray}
H_{MF} & = &  \sum_{k \sigma }\left[ (E_k^A-\mu)f^{A\dagger}_{k \sigma }
f^A_{k\sigma}
+ (E_k^B-\mu) f^{B\dagger}_{k \sigma } f^B_{k\sigma}\right]\nonumber \\
 & + & 2 N_L N\lambda_0\left( r_0^2-{1 \over 2} \right)
\end{eqnarray}
where we have transformed to the bonding/antibonding representation for
the fermionic operators
\begin{eqnarray}
f^A_{k\sigma}&=&\frac{1}{\sqrt{2}}\left(c_{k\sigma,1}+c_{k\sigma,2}\right) \\
f^B_{k\sigma}&=&\frac{1}{\sqrt{2}}\left(c_{k\sigma,1}-c_{k\sigma,2}\right)
\end{eqnarray}
and the respective energies are given by $E_k^{A/B}= r_0^2
(\varepsilon_k \pm t^{\perp}(k))$. $\varepsilon_k$ is the bare
in-plane dispersion which comprises nearest (t) and next-nearest
neighbor ($t'=\gamma t$) hopping 
\begin{equation}
\varepsilon_k=-2t[\cos(k_x)+\cos(k_y)+2\gamma \cos(k_x)\cos(k_y)]
\end{equation}
and $N_L$ denotes the number of sites per plane.
Note that at this level the square of the mean-field value of the
slave-boson field $b_0$, $b_0^2=Nr_0^2=N\delta /2$,
multiplicatively reduces both the inter- and intralayer hopping.
Fig. 1 shows the bonding and antibonding band for selected cuts
through the Brillouin zone. According to our choice for the
interlayer hopping Eq. (\ref{tperp}) the splitting is largest at
the $(\pi,0)$ points and reduces to $t^{\perp}_0$ along the
zone diagonals.
\begin{figure}[h]
%h=here, t=top, b=bottom, p=separate figure page
\begin{center}\leavevmode
\includegraphics[width=0.8\linewidth]{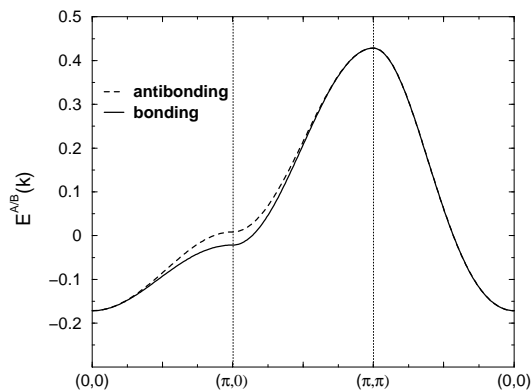}
\caption{Bonding and antibonding bands with respect to the Fermi
level for parameters $\gamma=-0.2$, $t^{\perp}_0=0.2$,
and doping $\delta=0.15$. Energies are measured in units of t.}
\end{center}
\end{figure}
As far as $\lambda_0$ is concerned, this quantity rigidly shifts the bare 
chemical potential $\mu_0$ as a function of doping and
is self-consistently determined by the following equation
\begin{eqnarray}
\lambda_0
= &-&\frac{1}{2N_L} \sum_k \left(\varepsilon_k + t^\perp(k)\right) f(E^A_k)
\nonumber \\
 &-&\frac{1}{2N_L} \sum_k \left(\varepsilon_k - t^\perp(k)\right) f(E^B_k)
\label{lambda}
\end{eqnarray}
where $f(E)$ is the Fermi function.

The presence of the coupling with the phonons introduces
new physical effects when one considers the fluctuations
of the bosonic fields. Since only a particular combination
$a=(A^{\dag}+A)/(2\sqrt{N})$ of the phonon fields $A$ and
$A^{\dag}$ is coupled to the fermions, it is more natural to
use the field $a$ and to integrate out the orthogonal
combination ${\widetilde{a}}=(A-A^{\dag})/(2\sqrt{N})$.
Then the quadratic action for the boson field $a$ reads
\begin{equation}
H_{\rm {phon}}=N\sum_{n,i,\alpha}
{{\omega_n^2+\omega_0^2}  \over {\omega_0}}a_{i,\alpha}^\dagger
a_{i,\alpha}\,\,, \label{aaction}
\end{equation}
where we have transformed the imaginary time into Matsubara
frequencies. Moreover, it is convenient to work in the
radial gauge \cite{RADGAUGE}, the
phase of the field  $b_{i,\alpha}=\sqrt{N}r_{i,\alpha}\exp(-i\phi) $
is gauged away
and only the modulus field $r_{i,\alpha}$ is kept, while $\lambda_{i,\alpha}$
acquires a time dependence $\lambda_{i,\alpha}\to \lambda_{i,\alpha}
+\partial_\tau \phi_{i,\alpha}$.
Thus one can define two three-component fields  ${\cal A}_{\alpha}^{\mu}
=(\delta r_{\alpha},\,\,\, \delta \lambda_{\alpha}, \,\,\, a_{\alpha})$
where  the time- and
space-dependent components are  the fluctuating part of the boson
fields $r_{i,\alpha} = r_0 \left( 1+\delta r_{i,\alpha}   \right)$,
$\lambda_{i,\alpha} = -i\lambda_0 + \delta \lambda_{i,\alpha}$ and
$a_{i,\alpha}$.

Writing the Hamiltonian of
coupled fermions and bosons as $H=H_{MF}+H_{\rm {bos}}+H_{\rm
{int}}$, where $H_{MF}$ is the above mean-field Hamiltonian,
which is quadratic in
the fermionic fields, $H_{\rm{bos}}$ is the purely bosonic part, also
including the terms with the $a$, $r$ and $\lambda$ bosons appearing
in the action (\ref{action}) and in $H_{\rm {phon}}$,
Eq.(\ref{aaction}). $H_{\rm{int}}$  contains the fermion-boson
interaction terms.
The inter- and intralayer hopping terms in the
bilayer $U=\infty$ Hubbard model give rise
to a leading order self-energy contribution
in the quadratic part of the bosonic Hamiltonian
\begin{eqnarray}
\label{selfenergy}
\Sigma^{intra}(q)  &  =  & \frac{r_0^2}{2N_L} \sum_k
 \varepsilon_{k-q} \left[f(E^A_k)+f(E^B_k)\right] \\
\Sigma^{inter}(q)  &  =  & \frac{r_0^2}{2N_L} \sum_k
 t^\perp(k-q) \left[f(E^A_k)-f(E^B_k)\right].\nonumber
\end{eqnarray}

For the following it is more convenient to transform also the bosonic
fields to symmetric and antisymmetric combinations respectively,
i.e. $\widetilde{\cal A}^{\mu}_{\pm}=\frac{1}{\sqrt{2}}\left({\cal A}_1^{\mu}
\pm {\cal A}_2^{\mu}\right)$ which we combine
into a single vector as
$\widetilde{\cal A}=(\widetilde{\cal A}_{+},\widetilde{\cal A}_{-})$.

After Fourier transformation to
momentum space, the bosonic part of the action reads
$$
H_{\rm {bos}}=N \sum_{q \mu \nu} \widetilde{\cal A}^{\mu}(q)B^{\mu \nu}(q)
\widetilde{\cal A}^{\nu}(-q)
$$
without explicitly indicating the frequency dependence for the
sake of simplicity and $\mu,\nu =r_+,\lambda_+ ,a_+,r_-,\lambda_- ,a_-$.
The matrix $B^{\mu,\nu}$, can be explicitely
determined from Eqs.(\ref{action})-(\ref{selfenergy}) which results in the
following block diagonal structure
\begin{equation}
\underline{\underline{\bf B}} = \left( \begin{array}{cc}
\underline{\underline{\bf B^+}} & {\bf 0} \\
{\bf 0}   & \underline{\underline{\bf B^-}}
\end{array} \right)
\end{equation}
and $\underline{\underline{\bf B^\pm}}$ denote the $3\times 3$ matrices
\begin{equation}
\underline{\underline{\bf B^\pm}}= \left( \begin{array}{ccc}
r_0^2\lambda_0 + \Sigma^{intra} \pm \Sigma^{inter}
& i r_0^2 & 0 \\
i r_0^2 & 0 & 0 \\
0 & 0 & \frac{\omega_n^2 + \omega_0^2}{\omega_0}
\end{array} \right).
\end{equation}

The last ingredients of our perturbation theory are the vertices
coupling the quasiparticles to the bosons. Similar to the bosonic
fields we combine them into two three-component vectors
$\widetilde{\bf \Lambda_{nm}}=\left(\widetilde{\bf \Lambda_{+,nm}},
\widetilde{\bf \Lambda_{-,nm}}\right)$
allowing us to write the interaction part of the Hamiltonian in the
form
\begin{equation}
H_{\rm {int}} =
\frac{1}{2N_L}\sum_{k,q,\sigma}\sum_{nm\mu\nu}
f^{n,\dagger}_{k+{q\over 2}\sigma}
\widetilde{\Lambda}^\mu_{nm}\left(k,q\right)f^m_{k-{q\over 2}
\sigma}\widetilde{\cal A}^\mu\left( q \right).
\end{equation}
where the indices $(nm)$ label the bonding and antibonding band respectively.
It turns out that only the following vertices give rise to a non-vanishing
coupling
\begin{eqnarray}
\widetilde{\bf \Lambda_{+,nn}} &=&\left(\begin{array}{c}
E_{k+{q\over 2}}^n + E_{k-{q\over 2}}^n \\
i \\
-2g \end{array}\right) \nonumber \\
\widetilde{\bf \Lambda_{-,AB}} &=&\left(\begin{array}{c}
E_{k+{q\over 2}}^A + E_{k-{q\over 2}}^B \\
i \\
-2g \end{array}\right) \nonumber \\
\widetilde{\bf \Lambda_{-,BA}} &=&\left(\begin{array}{c}
E_{k-{q\over 2}}^A + E_{k+{q\over 2}}^B \\
i \\
-2g \end{array}\right) \label{vert}
\end{eqnarray}

We are now in the position to evaluate the self-energy
corrections to the boson-propagators
\begin{equation}
D^{\mu \nu}(q,\omega_m)  =
\langle \widetilde{\cal A}^\mu(q,\omega_m)
\widetilde{\cal A}^\nu(-q,-\omega_m)\rangle
\end{equation}
which can be obtained from Dyson's equation
\begin{equation}
\underline{\underline{\bf D}}=\underline{\underline{\bf D^0}} -
\underline{\underline{\bf D^0}}
\underline{\underline{\bf \Pi}}\underline{\underline{\bf
D}}\label{dyson}.
\end{equation}
The zero order boson propagator is
\begin{equation}\label{D0}
\underline{\underline{\bf D^0}}=\frac{1}{2N}
\underline{\underline{\bf B^{-1}}}
\end{equation}
so that
\begin{equation}\label{BOSPROP}
\underline{\underline{\bf D}}=\lbrack 2 N
\underline{\underline{\bf B}}+ \underline{\underline{\bf
\Pi}}\rbrack^{-1}.
\end{equation}
The factor 2 multiplying the boson matrix B arises from the fact
that the bosonic fields in the radial gauge are
real and $\underline{\underline{\bf \Pi}}$ are just fermionic
bubbles with insertion of quasiparticle-boson vertices
\begin{eqnarray}
\Pi^{\mu \nu}(q,\omega_m)&=&
\frac{N}{2 N_L}\sum_{k}\sum_{st} {
{f\left( E^s_{k+{q\over 2}} \right)
-f\left( E^t_{k-{q\over 2}} \right)} \over
{ E^s_{k+{q\over 2}} - E^t_{k-{q\over 2}} -i \omega_m}} \nonumber \\
&\times& {\Lambda}_{st}^\mu\left( k,q \right)
{\Lambda}_{ts}^\nu\left( k,-q \right). \label{pi}
\end{eqnarray}
From the structure of the vertices Eqs. (\ref{vert}) one can see
that also $\underline{\underline{\bf \Pi}}$ acquires a block
diagonal structure
\begin{equation}
\underline{\underline{\bf \Pi}} = \left( \begin{array}{cc}
\underline{\underline{\bf \Pi^+}} & {\bf 0} \\
{\bf 0}   & \underline{\underline{\bf \Pi^-}}
\end{array} \right)
\end{equation}
with $\underline{\underline{\bf \Pi^\pm}}$ being symmetric
$3\times 3$ matrices.

Possible charge instabilities of the system can be deduced from
divergencies in the corresponding correlation functions
or scattering amplitudes.~\cite{ABRIK}
The above formal scheme allows to calculate the leading-order
expressions of the scattering amplitude both in the
particle-hole channel
\begin{equation}
\Gamma_{nm;st} (k,k';q,\omega)= -{1 \over 2}\sum_{\mu \nu}
 {\Lambda}_{nm}^{\mu}  \left(k', -q
\right) D^{\mu \nu} \left( q, \omega \right)
 {\Lambda}_{st}^{\nu} \left(k, q \right) \label{gamma}
\end{equation}
and in the particle-particle channel
\begin{eqnarray}
&& \Gamma_{nm;st}^C  (k,k';\omega) =
 - {1 \over 2} \sum_{\mu \nu}
 {\Lambda}_{nm}^{\mu}\left({k+k' \over 2} ,k'-k\right) \nonumber \\
&\times & D^{\mu \nu} (k-k',\omega)
 {\Lambda}_{st}^{\nu}\left(-{k+k' \over 2},k-k'\right)
\label{cooper}
\end{eqnarray}
It should be noted that the boson propagators are of order 1/N
(cf. Eq. (\ref{D0})) while the occurrence of a bare fermionic bubble leads to a
spin summation and is therefore associated with a factor N 
(cf. Eq. (\ref{pi})).
From Eq. (\ref{BOSPROP}) it thus follows that in this 1/N approach 
the quasiparticle scattering
amplitudes are residual interactions of order 1/N.

Since both the boson matrix $\underline{\underline{\bf B}}$ and
the polarizability matrix $\underline{\underline{\bf \Pi}}$ are
block diagonal the same also holds for the scattering amplitudes.
Moreover for the scattering of two quasiparticles on the Fermi
surface ($k=k_F; k'=k'_F$) one has only two different elements for 
the effective
scattering amplitude and we find in the particle-hole channel
\begin{eqnarray*}
\Gamma_S &\equiv& \Gamma_{AA;AA}=\Gamma_{AA;BB}=\Gamma_{BB;BB}
=\Gamma_{BB;AA} \\
\Gamma_A &\equiv& \Gamma_{AB;AB}=\Gamma_{BA;BA}=\Gamma_{AB;BA}
=\Gamma_{BA;AB}
\end{eqnarray*}
with
\begin{widetext}
\begin{equation}\label{gamsa}
\Gamma_{S/A}(k_F,k'_F,q,\omega)=
\frac{{B^{\pm}_{11}\over
\delta^2}-{g^2\over B_{33}}+{\Pi^{\pm}_{11}\over 4\delta^2} -
{2E_F\over \delta}-{E_F^2\over \delta^2}\Pi^{\pm}_{22}+i{E_F\over
\delta^2}\Pi^{\pm}_{12}}{1+\left({B^{\pm}_{11}\over
\delta^2}-{g^2\over
B_{33}}\right)\Pi^{\pm}_{22}-i{\Pi^{\pm}_{12}\over
\delta}+{\Pi^{\pm}_{11}\Pi^{\pm}_{22}-(\Pi^{\pm}_{12})^2\over
4\delta^2}}.
\end{equation}
\end{widetext}
For later use we also report here the evaluation of the
density-density response function which for the bilayer system
is denoted as
\begin{equation}
P_{st,nm}(q,i\omega)=\langle T \rho^{st}(q,i\omega)\rho^{nm}(q,-i\omega)\rangle
\end{equation}
where $\rho^{st}(q)=\sum_{k,\sigma}f_{k+q,\sigma}^{n\dagger}f_{k,\sigma}^m$ and
the indices $\{s,t,n,m\}=\{A,B\}$ refer to bonding and antibonding band states,
respectively.
The density-density response is most conveniently expressed via 
the particle-hole scattering amplitudes which yields
\begin{widetext}
\begin{equation}
P_{st,nm}(q,i\omega)=P^0_{st,nm}(q,i\omega)
-\frac{N}{N_L}\sum_{kk'}\frac{f(E_{k+{q\over 2}}^t)-f(E_{k-{q\over 2}}^s)}
{E_{k+{q\over 2}}^t - E_{k-{q\over 2}}^s-i\omega}
\Gamma_{mn,ts}^\pm(k,k',q,i\omega) 
\frac{f(E_{k'+{q\over 2}}^n)-f(E_{k'-{q\over 2}}^m)}
{E_{k'+{q\over 2}}^n - E_{k'-{q\over 2}}^m-i\omega}\label{PSR}
\end{equation}
\end{widetext}
The explicit form of the zero-order bubbles $P^0_{st,nm}(q,i\omega)$ 
is reported in Eqs. (\ref{P0EXP},\ref{P0MAT})
and from the block-diagonal structure of the scattering amplitude
it follows that also the elements of
$P_{st,nm}(q,i\omega)$ decouple into the symmetric and antisymmetric sector.

Additionally we report an approximate derivation for the 
scattering amplitudes in appendix \ref{apa} which neglects the 
k-dependence of the vertices.
This approach provides a more direct insight in the basic physical
aspects of the problem and we refer to it in the following
where appropriate.

\section{Results}
\subsection{Phase Separation}
Let us first consider the system without long-range forces. 
In this case our model is characterized by strong on-site
correlations which enable the electron-lattice coupling to drive
the system towards a phase separation instability due to the
strong reduction of the kinetic energy of the charge carriers.
In the bilayer system the occurrence of a phase separation
instability is signalled by a diverging scattering amplitude
$\Gamma_S$ in the symmetric sector (cf. Eq. (\ref{gamsa}))
in the limit $\omega=0$, $q\to 0$.
Since for zero temperature and $q\to 0$ the numerator of the 
fermionic bubbles  Eq. (\ref{pi}) in the symmetric sector corresponds
to a delta-function ($f\left( E^s_{k+0} \right)
-f\left( E^s_{k-0}\right)\sim\delta(E^s_k-\mu)$) the sum over
k-states only picks up contributions at the Fermi energy and as
a consequence the approximative scheme described in appendix \ref{apa} 
becomes exact.
Thus the instability criterion for PS in the bilayer system follows
from Eqs. (\ref{vintra},\ref{vinter}) and reads as
\begin{equation}\label{zero1}
1+\Gamma_S^0(0)\left[N_A(0)+N_B(0)\right]=0 .
\end{equation}
Here $N_{A/B}(0)$ refer to the density of states of bonding and antibonding
band at the chemical potential and $\Gamma_S^0$ is defined in Eq. 
(\ref{gams}). 
It is interesting to observe that to lowest order
there is no influence of $t^\perp$ onto the PS instability.
Since $\Sigma^{inter} < 0$ one could conclude from
Eq. (\ref{gams}) that interlayer charge fluctuations work cooperatively
with the electron-phonon interaction and support the PS instability.
However, interlayer hopping simultaneously enhances the kinetic energy
which reflects in an enhancement of the Lagrange parameter $\lambda_0$
(cf. Eq. (\ref{lambda})). In fact, the self-energy contributions cancel
out the $\lambda_0$ term in the 2nd order scattering amplitude
$\Gamma_S^0$ which reads as $\Gamma_S^0(\omega=0,q\to 0)=
\frac{1}{N}[-2E_F/\delta-g^2/\omega_0]$ and 
only indirectly depends on the interlayer hopping via
the Fermi energy.
Here the term proportional to the Fermi energy ($E_F<0$) corresponds to the
residual repulsion between the quasiparticles on the Fermi surface.
Despite the fact that we started from a infinite on-site repulsion
between bare particles the large screening in the system gives
rise to a finite scattering amplitude $\Gamma_S$.
Therefore the additional el.-ph. coupling can always turn the interaction
into an attractive one and eventually drive the system towards
phase separation.

In order to analyze in more detail the $q=0$
instabilities in the symmetric and antisymmetric sector it is
instructive to modify the model in Eq. (\ref{HHHam}) by coupling the
phonons to the full electron density rather than to density
fluctuations. In this way an el.-ph. coupling is effective already at
the mean-field level, where a nonzero mean-field value of the
phonon field arises in both symmetric ($a^0_+=\langle a_+\rangle$)
and antisymmetric ($a_-^0=\langle a_-\rangle$) combinations. It is
then straightforward to show that $\langle a_+\rangle$ induces a
doping dependent correction to the chemical potential and Eq.
(\ref{zero1}) is identical to the condition of a stationary point
in $\mu(\delta)$, signaling a divergence in the compressibility
$\kappa=-\partial\delta/\partial\mu$, i.e. phase separation (cf. appendix
in Ref. \onlinecite{BECCA}).
On the
other hand the instability in the antisymmetric sector 
corresponds to a second order phase transition
where the order parameter $\langle a_-\rangle$ 
starts to acquire a finite value. As a
consequence of different distortions in the two planes also the
corresponding charge densities will be different. 

\begin{figure}[htp]
%h=here, t=top, b=bottom, p=separate figure page
\includegraphics[width=8cm,clip=true]{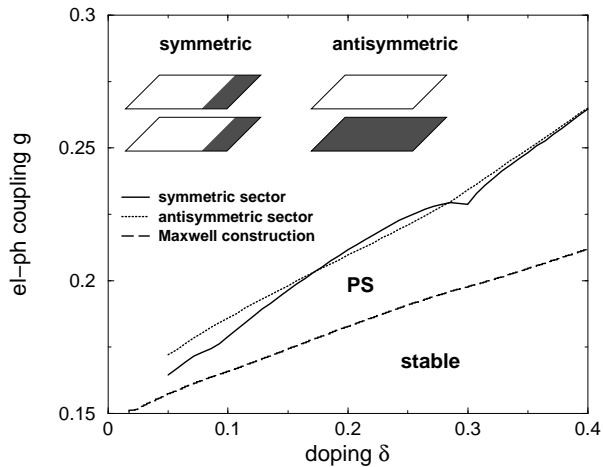}
\caption{Phase diagram electron-phonon coupling $g$ versus doping
$\delta$ for $q=0$ instabilities. The Maxwell construction for the 
(symmetric) phase transition separates the stable from the phase separated
region.
Parameters: $t=$1eV, $\gamma=-0.2$,
$t^0_{\perp}/t=0.2$. The inset sketches the symmetry-broken
state within the symmetric (i.e. PS) and antisymmetric channel.
Shaded areas indicate enhanced charge density.}
\end{figure}

Fig. 2 displays the phase diagram el.-ph. coupling
$g$ versus doping $\delta$ together with a sketch to elucidate the
symmetric and antisymmetric $q=0$ instabilities.  
For the PS instability line the van-Hove singularities of
bonding (BB) and antibonding band (AB) reflect as the two kinks at
$\delta\approx 0.1$ and $\delta\approx 0.3$, respectively.
The concentration range in between is characterized by 
an AB Fermi surface (FS) centered around $\Gamma=(0,0)$ 
and a BB FS centered around $X=(\pi,\pi)$. 
Since evaluation of the fermionic
bubbles in the antisymmetric sector Eq. (\ref{pi}) requires the summation over
an area which is determined by the difference of BB and AB 
Fermi surfaces, $\Gamma_A$ is strongly enhanced for $0.1 < \delta < 0.3$.  
It is therefore within this concentration range where
the antisymmetric 2nd order phase transition occurs before the
phase separation instability. However, a Maxwell construction has
to be done in order to properly determine the coexistence region
in the symmetric sector. The corresponding phase boundary is shown
by the dashed line in Fig. 2. From this Maxwell construction we
thus conclude that also in a bilayer system which is strongly
susceptible to a 2nd order instability in the antisymmetric sector
the presence of strong correlations favors the transition towards
a phase separated regime.

In principle our previous analysis does not exclude the occurrence of
a nesting induced phase transition before the PS instability line
is reached. In order to demonstrate that the instability really takes 
place at wave vector $q=0$ we report in
Fig. 3 the static scattering amplitudes in the
particle-hole channel $\Gamma_{S/A}(k_F,k'_F,q,\omega=0)$ obtained
from Eq. (\ref{gamsa}) for a fixed doping $\delta=0.25$ and tuning
the el.-ph. coupling towards the instabilities. In the absence of
el.-ph. coupling ($g=0$) the
interaction between quasiparticles is repulsive (i.e. $\Gamma>0$) in both 
the symmetric and antisymmetric channel.
Due to the strong local correlations this residual repulsion increases
with increasing wave vector $q$ in both channels. Upon switching on the
el.-ph. coupling $\Gamma_{S/A}(k_F,k'_F,q,\omega=0)$ therefore
becomes attractive for small $q$ and diverges at some
critical value $g_{crit}$.  
Note that results in Fig. 3 are evaluated for doping $\delta=0.25$
where the instability in the
antisymmetric sector occurs first. For this reason
$\Gamma_{A}$ becomes more negative for $q \rightarrow 0$ than
$\Gamma_{S}$.
\begin{figure}[h]
%h=here, t=top, b=bottom, p=separate figure page
\begin{center}\leavevmode
\includegraphics[width=7.5cm,clip=true]{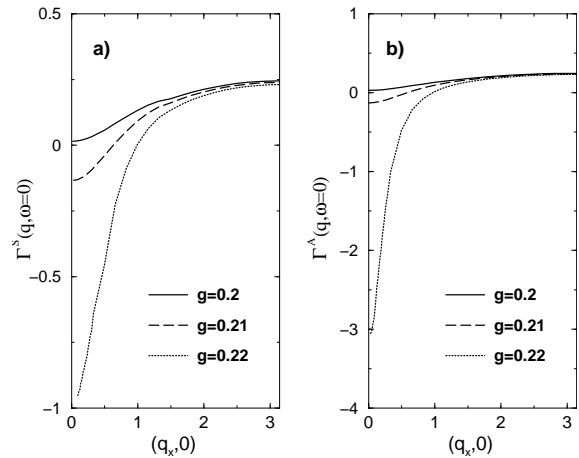}
\caption{Scattering amplitudes in the symmetric (a) and
antisymmetric (b) channel for a scan along $(q_x,0)$. The three
curves correspond to el.-ph. couplings $g=0.2,0.21,0.22$
respectively. Parameters: $\gamma=-0.2$,
$t^0_{\perp}/t=0.2$, doping $\delta=0.25$.}
\end{center}
\end{figure}

Let us now turn to the dynamical properties in the absence of
long-range interactions. Since we consider a strongly correlated
bilayer system the
zero sound mode consists of two branches. The acoustic one 
in the symmetric sector disperses as $\omega_s=v_F\sqrt{1+\tilde u}q$
for a 2-d system and an effective interaction $\tilde u$.~\cite{STERN}
A second, optical branch exists in the antisymmetric sector  
and its energy scale is determined by the interlayer hopping
$\omega_a\sim t^\perp$.
As a consequence the Boson propagator Eq. (\ref{BOSPROP}) has two
poles for each symmetry: 
one is the phonon and the other is the zero sound (which in
the symmetric sector becomes the 2-d plasmon when long-range
Coulomb forces  are included).
The two modes would cross each other (at rather small $q\sim \omega_0/v_F$)
if the phonon were decoupled
from the fermions but repel when the coupling is switched on.
Consequently one observes two spectral features in
each channel. For the symmetric combination (and small $q$): 
(a) the zero sound at low energy which upon increasing $g$ is pushed down and, 
becoming softer and softer, 
drives the PS; (b) the phonon mode at energy higher than $\omega_0$  
which is hardened since it is pushed up by the
repulsion with the zero sound.  
For the antisymmetric channel (a) at low momenta and small energies
appears the phonon which now upon increasing $g$ softens towards the 2nd order
instability and (b) at higher frequencies the zero sound optical
mode which for larger momenta is rapidly shifted to higher
frequencies and loses intensity. In
Fig. 4 we show the dispersion of these excitations along
the $(1,0)$ direction which can be obtained from the poles of
$\Gamma_{S/A}(k_F,k'_F,q,\omega)$ in Eq. (\ref{gamsa}).

\begin{figure}[h]
%h=here, t=top, b=bottom, p=separate figure page
\begin{center}\leavevmode
\includegraphics[width=0.7\linewidth]{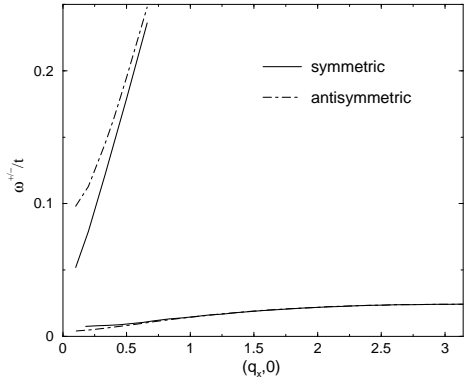}
\caption{Phonon and zero sound dispersions along the $(q_x,0)$ direction
in the symmetric ($\omega^+$,solid) and antisymmetric ($\omega^-$, dot-dashed)
channel.
Parameters: $\gamma=-0.2$, $t^0_{\perp}/t=0.2$, 
$g=0.22$, $\omega_{0}=0.04$,doping $\delta=0.25$.}
\end{center}
\end{figure}
\begin{figure}[htp]
%h=here, t=top, b=bottom, p=separate figure page
\includegraphics[width=6cm,clip=true]{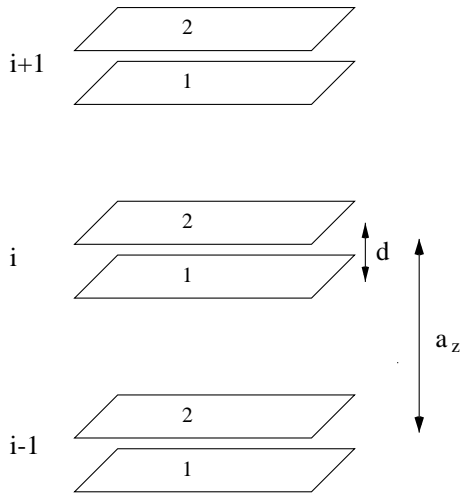}
\caption{Sketch of the bilayer structure which is used for
evaluation of the inter- and intralayer Coulomb interaction}
\end{figure}

\subsection{Inclusion of long-range interactions}
The nature of possible instabilities in the system (i.e.
phase separation or finite-q charge instabilities) 
crucially depends on the structure
of the long-range Coulomb (LRC) potential in the bilayer system.
Although we expect the effect of LRC forces to be most effective
in the small momentum-transfer case, where the underlying lattice
structure is less visible, we  explicitely
take into account the real symmetry of the bilayer square-lattice system
(cf. Fig. 5).

The intra- and interlayer contributions to the Coulomb potential
are derived in appendix \ref{AP2} and give rise to the following
interaction part of the Hamiltonian

\begin{equation}
H_C = \frac{1}{2N}\sum_{q,\mu,\nu=1,2}V_{\bf
q}^{\mu\nu}\rho_q^{\mu} \rho_q^{\nu}.
\end{equation}
where
\begin{eqnarray}
V_{\bf q}^{\mu=\nu}&=&V^{intra}_{{\bf q}_\parallel} (z=0)\nonumber \\
&=& -{V_C\over 2 }
\frac{A(q_x,q_y)}{\sqrt{\left[\frac{A^2(q_x,q_y)-1}{2\kappa(1-\kappa)}+1
\right]^2-1}}\label{vvintra}\\
V_{\bf q}^{\mu\neq\nu}&=&V^{inter}_{{\bf q}_\parallel} (z=0)\nonumber \\
&=& {V_C\over 2} \left\lbrace\frac{1+{1 \over 2} \frac{1}{1-\kappa}
[A^2(q_x,q_y)-1]}
{\sqrt{\left[\frac{A^2(q_x,q_y)-1}{2\kappa(1-\kappa)}+1
\right]^2-1}} -\kappa \right\rbrace \label{vvinter} .
\end{eqnarray}
The Coulombic coupling constant $V_C=e^2
a_z/(2\epsilon_{\perp}a_{xy}^2)$ has to range from roughly 0.5-3eV
in order to have holes in neighboring CuO$_2$ cells repelling each
other with a strength of 0.1-0.6 eV. Concerning the lattice
parameters typical values  in case of YBCO are $\kappa=d/a_z\approx 0.36$ and
$a_{xy}/a_z\approx 0.32$.
Note that $V_{\bf q}^{\mu\nu}$ is the potential between
electrons in a two-dimensional bilayer lattice and the small
momentum behavior reads as
\begin{eqnarray}
V^{intra}_{\bf q}&=&\frac{V_C}{\sqrt{8
\tilde{\epsilon}a_z^2}}{1 \over q} \label{vcintra} \\
V^{inter}_{\bf q}&=&V^{intra}_{\bf q}-\kappa \frac{V_C}
{2}.\label{vcinter}
\end{eqnarray}

Upon transforming the Coulomb potential to the (anti)symmetric representation 
\begin{equation}\label{vcsa}
V_{S/A}(q)={1 \over {2}} \left[ V^{intra}_{Coul}(q) \pm
V^{inter}_{Coul}(q)\right]
\end{equation}
it thus turns out that
the Coulombic contribution
approaches a constant value ($\approx 0.1 ... 0.5 eV$ depending on
parameters) in the antisymmetric sector for small momentum
transfer whereas the $q=0$
divergence in the symmetric part of the interaction naturally
leads to a suppression of phase separation as we will demonstrate 
in the next section.

\begin{figure}[htp]
%h=here, t=top, b=bottom, p=separate figure page
\includegraphics[width=8cm,clip=true]{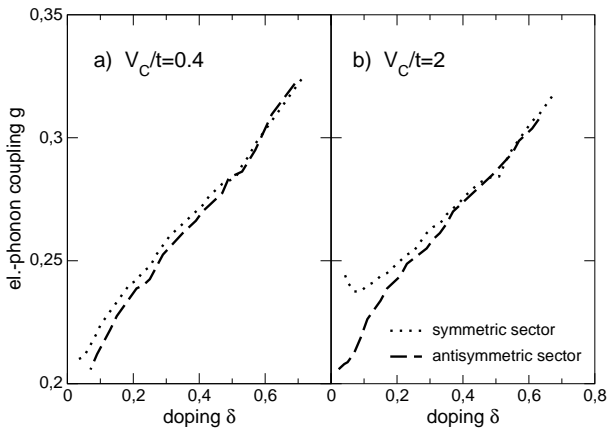}
\caption{ICDW instability lines in the phase diagram 
electron-phonon coupling $g$ versus doping
$\delta$. a) $V_C/t=0.4$; b) $V_C/t=2$. Parameters: $\gamma=-0.4$,
$\omega_0/t=0.04$, $t^0_{\perp}/t=0.2$,$V_C/t=2$,$\tilde{\epsilon}=6$,
$d/a_z=0.5$, $a_{xy}/a_z=0.32$.}
\end{figure}
\subsection{Analysis of CDW instabilities}
The analysis of possible instabilities under the presence
of long-range interactions is most conveniently carried out
by calculating the density-density response function $P^{LR}(q,\omega)$.
Using the corresponding short-range part $P^{LR}(q,\omega)$ Eq. (\ref{PSR})
one can perform the following resummation
\begin{equation}\label{RESUM}
\underline{\underline{\bf P^{LR}}}=\underline{\underline{\bf P^{SR}}}
+ \underline{\underline{\bf P^{SR}}}\,\,
\underline{\underline{\bf V^{Coul}}}\,\,\underline{\underline{\bf P^{LR}}}
\end{equation}
where the matrix $\underline{\underline{\bf V^{Coul}}}$ is given by
\begin{equation}
\underline{\underline{\bf V^{Coul}}} = \left(\begin{array}{cccc}
V^S(q) & V^S(q) & 0 & 0 \\
V^S(q) & V^S(q) & 0 & 0 \\
0 & 0 & V^A(q) & V^{A}(q) \\
0 & 0 & V^A(q) & V^A(q) \end{array}\right).
\end{equation}
Inverting Eq. (\ref{RESUM}) yields 
\begin{equation}
\underline{\underline{\bf P^{LR}}}=\lbrack \underline{\underline{\bf 1}}
-\underline{\underline{\bf P}}\,\, \underline{\underline{V^{Coul}}}\rbrack^{-1}
\underline{\underline{\bf P^{SR}}}
\end{equation}
so that the instabilities can be obtained from
\begin{widetext}
\begin{equation}\label{CDWCOND}
DET(\underline{\underline{\bf 1}}
-\underline{\underline{\bf P V^{Coul}}})=0=\left\{
\begin{array}{l}
1-V^S(q)\lbrack P_{AA,AA}^{SR}(q,0)+P_{AA,BB}^{SR}(q,0)+P_{BB,AA}^{SR}(q,0)+
P_{BB,BB}^{SR}(q,0)\rbrack \\
1-V^A(q)\lbrack P_{AB,AB}^{SR}(q,0)+P_{AB,BA}^{SR}(q,0)+P_{BA,AB}^{SR}(q,0)+
P_{BA,BA}^{SR}(q,0)\rbrack
\end{array}\right.
\end{equation}
\end{widetext}
for the symmetric and antisymmetric channel respectively.
First it should be noted that a diverging short-range density-density
response no longer results in a $q=0$ instability under the
presence of long-range interactions.
This is particularly obvious in the symmetric channel where the Coulomb 
potential behaves as $V^{S}(q\to 0) \to \infty$. Thus  one always finds 
a vanishing compressibility and phase separation is now ruled out.
However, in the antisymmetric channel a diverging short-range response
$P^{SR}_{n\neq m}(q\to 0,0)\to -\infty$ leads to a finite value of the
corresponding long-range polarizability
$P^{LR}_{n\neq m}(q=0,0) \sim -1/V^A(q=0)=-2a_z/(d V_c)$ according 
to Eqs. (\ref{vcintra}-\ref{vcsa}).  
Thus in this case LRC interactions 
leave the system to some
extend susceptible to long-wavelength antisymmetric density
fluctuations.

Concerning possible instabilities under the presence of
long-range forces the conditions in Eq. (\ref{CDWCOND}) can 
be in principle fulfilled inside the (short-range) instability regions of
symmetric and antisymmetric channel where $P^{SR}_{nm}(q,0)$ has a
positive branch up to some finite wave vector.
Due to the fact that $P^{LR}_{n\neq m}(q=0,0)$ stays finite 
the antisymmetric CDW instability can even
occur at arbitrarily small wave vectors depending on
the strength of the LRC interaction.

Fig. 6 depicts the phase diagram el.-ph. coupling versus doping
for two values of the  Coulomb interaction.
%Consider first  case of large LRC forces displayed in Fig. 6a.
Up to doping  $\delta \approx 0.4 ... 0.5$ 
the antisymmetric instability occurs at smaller
coupling $g$ than the symmetric one
whereas both instability lines merge for larger concentrations.
This behavior is best understood within the approximate formalism
given in the appendix. The corresponding RPA equations for the
effective interactions Eqs. (\ref{vintra},\ref{vinter}) suggest
that instabilities are favored for those wave-vectors $q_{crit}$ 
and dopings where (a) the residual quasiparticle 
interactions $\Gamma^0_S$ and $\Gamma^0_A$ Eqs.  (\ref{gams},\ref{gama})
have a minimum and (b) the 'bare' charge-charge correlations
Eq. (\ref{P0EXP}) are enhanced.
Concerning (a) Fig. 8 in appendix \ref{apa} shows a plot of 
$\Gamma^0_S$ and $\Gamma^0_A$ 
for the same values of $V_C$ also used in the results of Fig. 6.
The minima in these curves are determined by the relative strength
of Coulomb interaction and the residual repulsion due to the
slave-bosons. Since the latter part decreases with $\delta$ the
minima of $\Gamma^0_S$ and $\Gamma^0_A$ shift to larger $q_{crit}$
when doping is increased (cf. inset to Fig. 8 in
appendix \ref{apa}).
Moreover, in the limit $q\to 0$ the Coulomb interaction Eq. (\ref{vvinter})
approaches a constant in the antisymmetric sector and thus the 
minimum in  $\Gamma^0_A$ shifts to rather
low momenta when $V_C$ becomes sufficiently small (see the
corresponding discussion in appendix \ref{apa}).

The value of $q_{crit}$ is not only influenced by the structure
of $\Gamma^0_{S/A}$ but also by the 'bare'
charge-charge correlation functions as mentioned above.
For sizeable next-nearest neighbor hopping $t'$ those are naturally
enhanced for scattering processes connecting the high-density
sections of the (open) Fermi surface around $q=(\pi,0)$.
Whereas in the symmetric sector these processes are
between particle-hole (ph)states of the {\it same} Fermi surface,
the scattering is between {\it different} (i.e. bonding and antibonding)
Fermi surfaces in the antisymmetric sector implying a smaller
critical wave-vector $q_{crit}$ in the latter case.

Summarizing, the behavior of both $\Gamma^0_{S,A}$ and the bubbles 
Eq. (\ref{P0EXP}) indicates that the  
critical wave-vector $q_{crit.}$ is increasing with doping.
Additionally $q_{crit.}$ is expected to be smaller in the
antisymmetric sector (where it eventually tends to zero for
small $V_C$) than in the symmetric one. 
Therefore, at small $\delta$ (corresponding to a 
small critical wave-vector 
$q_{crit.}$) the electron-phonon coupling has to overcome the large
$1/q_{crit}$ Coulomb repulsion in the symmetric channel.   
However, in the antisymmetric sector the Coulomb repulsion
approaches a constant in the limit $q \to 0$ so that the CDW instability
is reached for much smaller $g$ in this case. 
At larger doping  $q_{crit.}$ is shifted to larger
values and consequently the critical couplings $g_{crit}$ 
in both channels approximately coincide due to the
vanishing difference between $V^S(q)$ and $V^A(q)$ for larger wave-vectors.

Finally, Fig. 7 displays the density-density 
response functions in both channels
for $\delta=0.1$ close to the 
respective instabilities.
Near the instability line the system is characterized by a significant 
quasiparticle attraction within a large portion of momentum space.
The orientation of the critical wave-vectors is strongly determined
by the structure of the bubbles  Eq. (\ref{P0EXP}). For our
choice of $t'/t=-0.4$ (as appropriate for Bi2212) 
those are naturally enhanced along the $(\pi,0)$
axis of the Brillouin zone favoring singular scattering in the same
direction for both channels.  

\begin{figure}[htp]
%h=here, t=top, b=bottom, p=separate figure page
\includegraphics[width=8cm,clip=true]{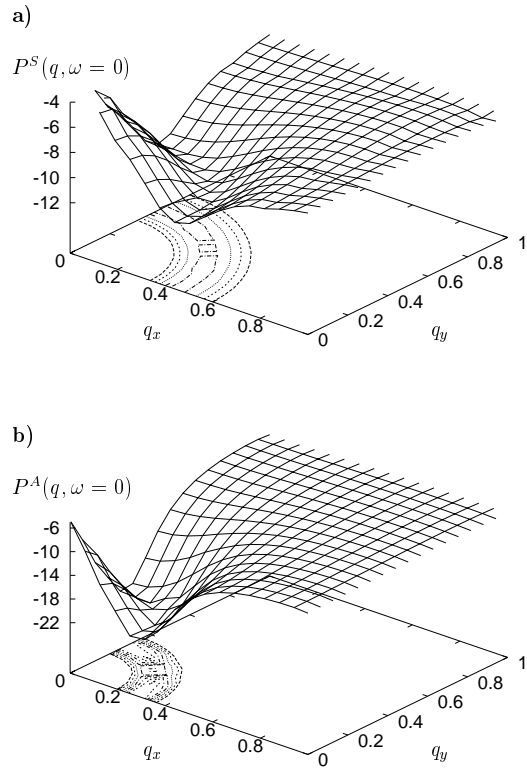}
\caption{Density-density response function in the symmetric (a)
and antisymmetric (b) sector 
close to the ICDW instability. Parameters: $\gamma=-0.4$,
$g/t=0.23$, $\delta=0.1$, $\omega_0/t=0.04$, 
$t^0_{\perp}/t=0.2$,$V_C/t=2$,$\tilde{\epsilon}=6$,
$d/a_z=0.5$, $a_{xy}/a_z=0.32$.}
\end{figure}

\section{Discussion and conclusions}
\subsection{Influence on the superconducting gap structure 
in a bilayer system}
As shown in Sec. II the singular scattering amplitudes in the 
particle-hole channel which
occur close to the CDW instabilities are naturally connected with  singular
attractive interactions in the particle-particle channel.
In this section we briefly discuss the consequences for
the superconducting gap structure in a bilayer material based
on the results derived above. 
When we fix the electron-phonon coupling constant $g$ to some material
specific value Fig. 6 suggests the investigation of the two following cases.
\begin{enumerate}
\item For small $V_C$ and arbitrary $g$ (Fig. 6a) or large 
$V_C$ and large $g$ (Fig. 6b)
the instabilities in the symmetric and antisymmetric 
sectors occur almost at the same doping concentration. 
Therefore the associated critical fluctuations in both channels are comparable.
\item At small $g$ and large $V_C$ the pairing interaction occurs 
dominantly in the antisymmetric channel (Fig. 6b).
\end{enumerate}
Both cases can be treated within the two-band model
(cf. Suhl et al. \cite{SUHL}) when we restrict on the
static part of the effective interactions $\Gamma^{S/A}_q$ 
in the symmetric and antisymmetric channel.
A similar analysis for the
one band-model has been performed in Ref. \onlinecite{PERALI}.
The SC gap for bonding and antibonding band can be introduced via
\begin{eqnarray}
\!\!\!\!\!\!\!\!\!\Delta_k^{A}\!\!\!&=&\!\!\!-{1\over N}\sum_q\lbrack \Gamma^S_q \langle
f_{k+q,\uparrow}^{A\dagger}f_{-k-q,\downarrow}^{A\dagger}\rangle
+ \Gamma^A_q \langle
f_{k+q,\uparrow}^{B\dagger}f_{-k-q,\downarrow}^{B\dagger}\rangle\rbrack 
\label{DA}\,\,\,\,\,\\
\!\!\!\!\!\!\!\!\!\Delta_k^{B}\!\!\!&=&\!\!\!-{1\over N}\sum_q\lbrack \Gamma^A_q \langle
f_{k+q,\uparrow}^{A\dagger}f_{-k-q,\downarrow}^{A\dagger}\rangle
+ \Gamma^S_q \langle
f_{k+q,\uparrow}^{B\dagger}f_{-k-q,\downarrow}^{B\dagger}\rangle\rbrack
\label{DB}\,\,\,\,\,
\end{eqnarray}
and for simplicity we neglect interband pairing $\Delta_k^{AB} \approx 0$.
In order to derive some analytical results we further assume the
interaction $\Gamma^{S/A}$ to be constant within some energy
range $2\omega_c$ around E$_F$ \cite{NOTE2}.
Then the self-consistency equation for the bilayer system reads as
\begin{equation}\label{BCS2L}
\lbrack 1-\Gamma^S N_A F_A\rbrack \lbrack 1-\Gamma^S N_B F_B\rbrack
= (\Gamma^A)^2 N_A N_B F_A F_B
\end{equation}
where $N_{(A)B}$ denotes the DOS for the (anti)bonding band near $E_F$ and
$F_{A/B}(\Delta_{A/B})=\int_0^{\omega_c} {d\epsilon\over\sqrt{\epsilon^2
+\Delta_{A/B}^2}}th\lbrack \sqrt{\epsilon^2+\Delta_{A/B}^2}/ (2kT)\rbrack$.

Let us now consider the first case 
where the symmetric and antisymmetric instability lines
are close so that both $\Gamma^A$ and $\Gamma^S$ display
singular behavior near some critical doping, i.e.
$\Gamma^A \approx \Gamma^S \to \Gamma^{intra}/2$.
As a consequence Eq. (\ref{BCS2L}) simplifies to 
$1=\Gamma^{intra}F(0)(N_A+N_B)/2$ when we are at the
transition temperature which therefore yields
\begin{equation}
kT_c=1.14\omega_c \exp(-1/(\Gamma^{intra}{N_A+N_B\over 2})).
\end{equation}
Moreover, below T$_c$ we have the same gap value in
bonding and antibonding band $\Delta_A=\Delta_B$.
Close to the instability line the layers appear to be 
decoupled with respect to the pairing interaction which is 
almost completely due to intralayer scattering.
Thus the situation in this case is equivalent to the
single layer model investigated in Ref. \onlinecite{PERALI}
with an effective DOS ${N_A+N_B\over 2}$.
On the other hand in the second case where
symmetric and antisymmetric instability lines are well separated
only $\Gamma^A$ becomes singular near the critical doping.
Note that in addition to a negatively diverging in-plane
response this implies a large repulsive inter-plane interaction
as can be deduced from Eqs. (\ref{vintra},\ref{vinter}).
In this case Eq. (\ref{BCS2L}) reduces to 
$1= (\Gamma^A)^2 N_A N_B F_A F_B$ and the transition
temperature is obtained as 
\begin{equation}
kT_c=1.14\omega_c \exp(-1/(\Gamma^{A}\sqrt{N_A N_B})).
\end{equation}
Below T$_c$ we have now $\Delta_A \neq \Delta_B$ and
it turns out from Eqs. (\ref{DA},\ref{DB}) that 
the superconducting gap in the antibonding band
is determined by the pair correlations of the bonding band
and vice versa.

In optimally doped bilayer cuprates the van-Hove singularity 
of the AB is quite close to the Fermi level. Therefore one should
expect that around $\delta_{opt}$ the pair correlations in the
AB exceed those of the BB, i.e. 
$\langle f_{k,\uparrow}^{A\dagger}f_{-k,\downarrow}^{A\dagger}\rangle
> \langle f_{k,\uparrow}^{B\dagger}f_{-k,\downarrow}^{B\dagger}\rangle$.
This implies that around optimal doping
$\Delta_B > \Delta_A$ when the scattering is dominantly
antisymmetric 
but $\Delta_B = \Delta_A$ when the pairing interaction  takes
place in the symmetric channel.
Up to now the energy gaps below T$_c$ for both antibonding and bonding band
have been examined in detail only for an underdoped
modulation-free Pb-Bi2212 sample in Ref. \onlinecite{BORIS}.
It turns out that both gaps are identical and deviate significantly
from d-wave symmetry around the nodal direction.
This may be associated with a normal state contribution to the
gap (pseudogap) in this sample since a preliminary analysis revealed a similar
anisotropy above T$_c$.
Within our analysis two identical energy gaps would imply an underlying
interaction where symmetric and antisymmetric components are of
similar strength. However, pair correlations can only be significantly 
influenced by the van Hove singularity when 
it is separated from $E_F$  within an energy scale of $T_c$.
This is probably not the case for this
particular underdoped sample. In this regard it would be interesting to   
repeat the same analysis for an optimally doped sample .

\subsection{Influence on the normal state resistivity}
Our last point concerns the result that in the antisymmetric sector 
close to the instability singular fluctuations with $q \approx 0$
are possible. In fact this feature is not restricted to bilayer
materials but should also occur in single layer compounds when
one includes the Coulomb interaction between individual layers.
Denoting the in-plane momenta with $q_{||}$ and the perpendicular
momentum with $q_\perp$ it is known \cite{BILL} that
the Coulomb potential diverges $ \sim 1/q_{||}^2$ for $q_{\perp}=0$.
For finite $q_{\perp}$ the potential $V_C(q_{||}=0)$ remains finite with the
smallest repulsion for $q_{\perp}=\pi$.
Within the same model investigated in this paper but extended to
a real three-dimensional layered structure the possibility arises for
singular in-plane fluctuations $q_{||}=0$ for $q_{\perp}=\pi$.
This on the other hand could have important consequences for
temperature dependent transport properties such as the electrical conductivity.
Based on a Boltzmann-equation approach Hlubina and Rice \cite{HLUB} have 
evaluated the resistivity for models which are
characterized by strong (critical) scattering between selected points on the
Fermi surface. The associated electron lifetime and resistivity 
displays an anomalous temperature dependence which, however, is
short-circuited by the remainding electrons on the rest of the
Fermi surface ('cold regions') which scattering rate is of the standard
Fermi liquid form $1/\tau \sim T^2$.
As a consequence it was found that
the resistivity has the standard Fermi
liquid form $\rho \sim T^2$ up to some energy scale which is
determined by the distance to the critical point.
In a model with singular scattering at low momenta as discussed above 
all points on the Fermi surface would correspond to 'hot spots', therefore
the short-circuit problem would be prevented and the critical scattering
would determine the temperature dependence of the resistivity
down to $T=0$.

\subsection{Conclusion}
We have investigated the possible charge instabilities of a
bilayer Hubbard-Holstein model. In particular we have focused
on the question wether these instabilities preferably occur in the symmetric
or antisymmetric channel with respect to the exchange of the layers.

In the absence of long-range Coulomb interactions and similar
to the single-layer case \cite{BECCA} 
our calculations support the existence of phase separation arising from
the attractive electron-phonon interaction.
However, both the symmetric and antisymmetric instability
lines are rather close (cf. Fig. 2) and we find that also the 
antisymmetric symmetry-breaking (which corresponds to a second order
phase transition) occurs at wave-vector $q=0$ (cf. Fig. 3b) due
to the strong on-site correlations. 
Hence in the bilayer model
phase separation is solely supported
due to the Maxwell construction which only applies in the
symmetric sector where the phase transition is first order.

Inclusion of long-range forces spoils phase separation
but finite-momentum instabilities  still take place
in the symmetric sector of the charge-charge correlations. 
In the antisymmetric sector the critical wave-vector crucially
depends on the strength of the long-range Coulomb interaction $V_c$ and
for sufficiently small $V_c$ and low doping can be still around
$q \approx 0$. 
Moreover, since both types of instabilities now correspond  to a 2nd order
phase transition the  Maxwell construction does not apply and
one finds that the antisymmetric instability is now favored especially
at low doping.

We have discussed the above findings in the context of  high-T$_c$ 
superconductors. Recent progress in the resolution of ARPES
experiments has made it possible to separably detect the superconducting
gaps in bonding and antibonding band respectively. We have argued that 
from the relative sizes of the gaps one can in principle deduce the
symmetry of the underlying interaction. Within the Hubbard-Holstein
model we find two possibilities, depending on the strength of
electron-phonon coupling $g$ and the long-range Coulomb interaction $V_c$.
Sizeable el.-ph. coupling $g\approx 0.3t$ (Fig. 6a,b) implies  that in 
the quantum 
critical region (which in the $g-\delta$ phase diagram is
at rather large doping) symmetric  and antisymmetric fluctuations 
are comparable.
On the other hand the scattering is dominantly antisymmetric in case of 
small $g$ and large $V_c$ (Fig. 6b) when one approaches the instability.
Moreover this antisymmetric transition now occurs at concentrations
$\delta \approx 0.1 ... 0.2$ (cf. Fig. 6b) which 
covers the range where T$_c$ is largest in the high-T$_c$ cuprates and
therefore is more compatible with the quantum critical point scenario.
Having in mind that antiferromagnetic fluctuations in the bilayer
high-T$_c$ cuprates are also antisymmetric with respect to exchangge
of the layers \cite{TRAN} both charge and spin fluctuations
may easily coexist and determine cooperatively 
the unusual properties of the cuprates.
Unfortunately our leading order analysis in $1/N$ 
of the $U\to \infty$ Hubbard-Holstein model only captures the
charge instabilities of the model whereas antiferromagnetic
correlations would appear at higher order in $1/N$.
Therefore a long way is still to be followed in order to
formalize the interplay between
charge and spin degrees of freedom and to answer the question 
how charge instabilities
are mirrored in the spin criticality. This intriguing but difficult
issue is definitely beyond the scope of the present paper
but should be definitely investigated in future work.

\acknowledgments
This work was supported by the Deutsche Forschungsgemeinschaft
under contract SE$806/6-1$. I have benefitted from 
interesting and valuable discussions on this issue with 
C. Di Castro and M. Grilli.
Also thanks to S. Varlamov for his assistance concerning the 
gap structure within the two-band model. 
Finally I acknowledge hospitality of the Dipartimento di Fisica of 
Universit\`a di Roma ``La Sapienza'' where part of this work has 
been completed.

\appendix
\section{Approximate evaluation of the instabilities} \label{apa}
For a first analysis of the instabilities where only the
$\omega=0$ behavior of the bubbles is relevant, it is convenient to
simplify the formalism of Sec. II by approximating the vertices in the
following way:
\begin{equation}\label{approxi}
{\Lambda}_{st}^\mu\left( k,q \right) \to
{\Lambda}_{st}^\mu\left( k_F,q \right)
\end{equation}
i.e. restricting the quasiparticle momenta to $k_F$.
As a consequence the structure of $\underline{\underline{\bf \Pi}}$
Eq. (\ref{pi}) simplifies to
\begin{displaymath}
\Pi^{\mu \nu}(q,\omega_m)=\frac{1}{2}{\Lambda}_{st}^\mu\left( k_F,q
\right){\Lambda}_{ts}^\nu\left( k_F,-q \right)P^0_{st;ts}
\end{displaymath}
where $P^0_{st;ts}$ are now the usual fermionic bubbles
\begin{equation}\label{P0EXP}
P^0_{st;ts}=\frac{N}{N_L}\sum_{k}
{{f\left( E^s_{k+{q\over 2}} \right)
-f\left( E^t_{k-{q\over 2}} \right)} \over
{ E^s_{k+{q\over 2}} - E^t_{k-{q\over 2}} -i \omega_m}}
\end{equation}
with $s=A,B$ and $t=A,B$ respectively.
The matrix representation of $P^0_{st;ts}$ 
thus acquires a block diagonal structure
\begin{equation}\label{P0MAT}
\underline{\underline{\bf P^0}} = \left(\begin{array}{cccc}
P^0_{AA;AA} & 0 & 0 & 0 \\
0 & P^0_{BB;BB} & 0 & 0 \\
0 & 0 & 0 & P^0_{AB;BA} \\
0 & 0 & P^0_{BA;AB} & 0 \end{array}\right).
\end{equation}

Eq. (\ref{dyson}) can now be rewritten as a Dyson equation for
the scattering amplitudes
\begin{equation}
\underline{\underline{\bf \Gamma}}=\underline{\underline{\bf \Gamma^0}}
+ \underline{\underline{\bf \Gamma^0}}
\underline{\underline{\bf P^0}}\underline{\underline{\bf \Gamma}}
\end{equation}
and the second order scattering amplitude
$\Gamma^0_{st,nm}$ is also block diagonal
\begin{equation}
\underline{\underline{\bf \Gamma^0}} = \left(\begin{array}{cccc}
\Gamma^0_{AA;AA} & \Gamma^0_{AA;BB} & 0 & 0 \\
\Gamma^0_{BB;AA} & \Gamma^0_{BB;BB} & 0 & 0 \\
0 & 0 & \Gamma^0_{AB;AB} & \Gamma^0_{AB;BA} \\
0 & 0 & \Gamma^0_{BA;AB} & \Gamma^0_{BA;BA} \end{array}\right).
\end{equation}
with only two different elements
\begin{eqnarray}
\Gamma^0_S &\equiv&
\Gamma^0_{AA;AA}=\Gamma^0_{AA;BB}=\Gamma^0_{BB;BB}
=\Gamma^0_{BB;AA} \label{gams} \\
&=&\frac{1}{N}\left[-\frac{E_F}{r_0^2}+\frac{\lambda_0}{4r_0^2}+\frac{\Sigma^{intra}
+\Sigma^{inter}}{4r_0^4}-\frac{g^2}{\omega_0}\right]\nonumber  \\
\Gamma^0_A &\equiv&
\Gamma^0_{AB;AB}=\Gamma^0_{BA;BA}=\Gamma^0_{AB;BA}
=\Gamma^0_{BA;AB} \label{gama} \\
&=&\frac{1}{N}\left[-\frac{E_F}{r_0^2}+\frac{\lambda_0}{4r_0^2}+\frac{\Sigma^{intra}
-\Sigma^{inter}}{4r_0^4}-\frac{g^2}{\omega_0}\right] \nonumber
\end{eqnarray}

\begin{figure}[htp]\label{figgam}
%h=here, t=top, b=bottom, p=separate figure page
\includegraphics[width=7.5cm,clip=true]{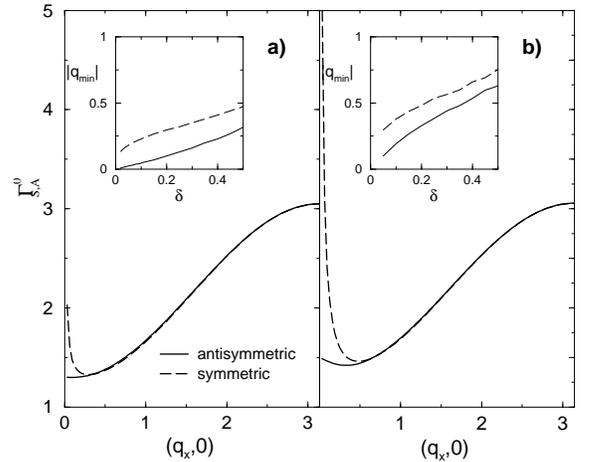}
\caption{Scattering amplitudes $\Gamma^0_{S,A}$ along the $(q_x,q_y=0)$
direction of the Brillouin zone for $V_c=0.4t$ (a) and $V_c=2t$ (b).
The electron-phonon interaction
leads to a constant shift of the curves which are plotted for $g=0$.
The insets show the doping dependent length of the wave-vector corresponding
to the minimum of $\Gamma^0_S$ and $\Gamma^0_A$ respectively. 
Parameters: $t=1eV$, $\gamma=-0.4$, $\delta=0.2$, 
$t^0_{\perp}/t=0.2$,$V_C/t=2$,$\tilde{\epsilon}=6$,
$d/a_z=0.5$, $a_{xy}/a_z=0.32$.}
\end{figure}

Within this framework long-range Coulomb interactions can be easily
incorporated by adding 
their symmetric and antisymmetric combinations Eqs. (\ref{vcsa}) 
to the scattering amplitudes Eqs. (\ref{gams},\ref{gama}).
Fig. \ref{figgam} displays $\Gamma^0_S$ and $\Gamma^0_A$ along
the $(1,0)$ direction of the Brillouin zone. The small $q$ behavior
is dominated by the LRC interaction which diverges as $1/q$ in
the symmetric channel but approaches a constant in case of the
antisymmetric potential. For large wave-vectors both curves merge
since in this regime the interaction is determined by
the residual repulsion mediated by the slave-bosons.

Finally, due to the block diagonal structure of both the scattering
amplitude and $P^0_{st;nm}$ the RPA problem decouples into two
$2\times 2$ matrix equations. 
The RPA scattering amplitudes for
intra- and interlayer scattering are given by
\begin{widetext}
\begin{eqnarray}
\Gamma^{intra}(q)&=&\frac{\Gamma_S^0}{1-\Gamma_S^0(P^0_{AA;AA}+P^0_{BB;BB})}
+ \frac{\Gamma_A^0}{1-\Gamma_A^0(P^0_{AB;BA}+P^0_{BA;AB})} \label{vintra}\\
\Gamma^{inter}(q)&=&\frac{\Gamma_S^0}{1-\Gamma_S^0(P^0_{AA;AA}+P^0_{BB;BB})}
- \frac{\Gamma_A^0}{1-\Gamma_A^0(P^0_{AB;BA}+P^0_{BA;AB})} \label{vinter}
\end{eqnarray}
\end{widetext}
For $\omega=0$ the fermionic bubbles only display a weak momentum dependence
up to the Fermi wave-vector $k_F$. Therefore the 
instability vectors $q_c$ as arising from Eqs. 
(\ref{vintra},\ref{vinter}) are approximately determined by 
the minimum $\Gamma^0_{S,A}(q_{min})$ of the scattering amplitudes 
Eqs. (\ref{gams},\ref{gama}).
The insets to Fig. 8 display the doping dependence of $|q_{min}|$
for two values of the Coulomb potential.
Within our 2-dimensional model the minimum $q_{min}$ 
in $\Gamma^0_A$ always occurs at finite (but arbitrarily small)
momenta since $V_C^A\sim C-Aq$ and the residual repulsion of the
slave-bosons behaves as $\sim q^2$. Note that a complete 
3-dimensional treatment would yield $V_C^A\sim C-Aq^2$
so that in this case a true $q=0$ instability could be realized.

\section{Derivation of the Coulomb potential in a bilayer system}\label{AP2}
In order to derive an explicit expression for the Coulomb potential
in the spirit of the point-charge approximation
we start from the discretized form of the Laplace equation.
Moreover, since we assume that our two-dimensional model
represents planes of a truly three-dimensional lattice
we also include a third spacial dimension. For clarity
we restore the explicit dependence of the
in-plane lattice spacing $a_{xy}$, which in Sec. II was
set to unity in the square two-dimensional lattice. In the
third space direction, instead, we assume the unit cell to
have a lattice spacing $a_z$. In addition each unit cell contains two layers
separated by spacing $d$.
Note that in the present case (with non-equidistant sampling points
along the z-direction) the second derivative of a function $f$
at point $x_i$ can be
represented as
\begin{eqnarray}
f''(x_i)&=&\frac{2}{h_ih_{i-1}(h_i+h_{i-1})}
\times\left[h_{i-1}f(x_{i+1})\right.\\
&+&\left. h_i f(x_i-1) -(h_i+h_{i-1})f(x_i)\right]\nonumber
\end{eqnarray}
where $h_i$ is the distance between points $x_i$ and $x_{i+1}$.

Due to the presence
of two layers in the unit cell one obtains the following two coupled Laplace
equations for the Coulomb potential $\phi$: 
\begin{eqnarray*}
-e \delta (R^2_i-R^2_j)&=&\frac{\epsilon_{\parallel}}{a_{xy}^2}\sum_{\eta=x,y}
\left[ \phi(R^2_i-R^2_j+\eta)+\right.\nonumber \\
&+& \left.\phi(R^2_i-R^2_j-\eta)-2 \phi(R^2_i-R^2_j)
\right]
+ \nonumber \\
&+& \frac{2\epsilon_{\perp}}{d a_z(a_z-d)}
\left[ d\phi(R^1_{i+1}-R^2_j)\right.\nonumber\\
&+&\left. (a_z-d)\phi(R^1_i-R^2_j)-
a_z \phi(R^2_i-R^2_j) \right]
 \nonumber \\
0&=&\frac{\epsilon_{\parallel}}{a_{xy}^2} \sum_{\eta=x,y}
\left[ \phi(R^1_i-R^2_j+\eta)+\right.\nonumber\\
&+&\left. \phi(R^1_i-R^2_j-\eta)-2 \phi(R^1_i-R^2_j)
\right] + \nonumber \\
&+&
\frac{2\epsilon_{\perp}}{d a_z(a_z-d)}
\left[ d\phi(R^2_{i-1}-R^2_j)\right.\nonumber\\
&+&\left. (a_z-d)\phi(R^2_i-R^2_j)-
a_z \phi(R^1_i-R^2_j)
 \right]
\end{eqnarray*}
where $R^{\alpha}_i$ denote lattice sites on plane $\alpha$ and
$\epsilon_{\perp}$ and
$\epsilon_{\parallel}$ are the high-frequency dielectric constants
perpendicularly  and along the planes respectively.
The corresponding Fourier transformed equations read as
\begin{eqnarray*}
-e\frac{a_z^2\kappa(1-\kappa)}{2\epsilon_{\perp}}=
A(q_x,q_y)\phi^{22}_{\bf q}
+ \left[\kappa \exp(iq_z a_z)+1-\kappa\right]
\phi^{12}_{\bf q}&& \\
0=A(q_x,q_y)\phi^{12}_{\bf q} + \left[\kappa \exp(-iq_z a_z)+1-\kappa\right]
\phi^{22}_{\bf q}&&
\end{eqnarray*}
with $\tilde{\epsilon}\equiv \epsilon_{\parallel}/
\epsilon_{\perp}$, $\kappa=d/a_z$ and the in-plane momentum dependence
is contained in
\begin{displaymath}
A(q_x,q_y) = \tilde{\epsilon}{\kappa(1-\kappa)\over (a_{xy}/a_z)^2 } \left[
\cos (a_{xy}q_x) + \cos (a_{xy}q_y) -2 \right] -1\,\,.
\end{displaymath}

We thus obtain for the
in- and intra-plane LRC potential in three-dimensional momentum space
\begin{eqnarray*}
\phi^{intra}_{\bf q} = -{e a_z^2 \over 8 \epsilon_{\perp}}
\frac{A(q_x,q_y)}{\frac{A^2(q_x,q_y)-1}{4\kappa(1-\kappa)}+
\sin^2(\frac{q_za_z}{2})} \\
\phi^{inter}_{\bf q} = {e a_z^2 \over 8 \epsilon_{\perp}}
\frac{\kappa\exp(-iq_xa_z)+1-\kappa}{\frac{A^2(q_x,q_y)-1}{4\kappa(1-\kappa)}+
\sin^2(\frac{q_za_z}{2})}.
\end{eqnarray*}
Since we are interested in the effects of the Coulomb potential
on the two-layer system, we now transform from
$q_z$ to real space for the $z=0$ unit cell  obtaining
\begin{eqnarray*}
\phi^{intra}_{{\bf q}_\parallel} (z=0)&=& -{e a_z \over 4
\epsilon_{\perp}}
\frac{A(q_x,q_y)}{\sqrt{\left[\frac{A^2(q_x,q_y)-1}{2\kappa(1-\kappa)}+1
\right]^2-1}}\\
\phi^{inter}_{{\bf q}_\parallel} (z=0)&=& {e a_z \over 4
\epsilon_{\perp}} \left\lbrace\frac{1+{1 \over 2}
\frac{1}{1-\kappa} [A^2(q_x,q_y)-1]}
{\sqrt{\left[\frac{A^2(q_x,q_y)-1}{2\kappa(1-\kappa)}+1
\right]^2-1}} -\kappa \right\rbrace
\end{eqnarray*}
In the limit $\kappa=1/2$ and $a_z \to 2d$ one recovers the result
of the single-layer calculation of F. Becca et al.\cite{BECCA} in which case
$\phi^{inter}_{{\bf q}}$ denotes the potential between successive
layers.


\begin{thebibliography}{9}
\bibitem{COMM} The electron-phonon coupling to density fluctuations can 
               be viewed as originating from a displacement transformation
               at every lattice site. In this way one can eliminate 
               phononic contributions to the
               ground state energy on the mean-field level.
\bibitem{NOTE2} The range of attraction of the effective interaction
                has in principle to be defined in q-space 
                (cf. Ref. \protect\onlinecite{PERALI}). It crucially
                depends on the distance to the instability line in the
                $(\delta,g)$ phase diagram. 
                When one stays sufficiently away from the quantum critical 
                line, attraction is restricted
                to wave-vectors up to $q_c$ and corrispondingly the 
                cutoff energy
                can be approximated as $\omega_c \approx v_F  q_c$ where $v_F$
                is the Fermi velocity. However, at the QCP the interaction 
                is attractive over the
                whole Brillouin zone and therefore $\omega_c$ is of the 
                order of the bandwidth.
\end{thebibliography}
\end{document}